\documentclass[aps,twocolumn,pra,superscriptaddress]{revtex4}
\usepackage{epsfig,graphicx,times}
\usepackage{amstext}
\usepackage{amsmath}
\usepackage{amssymb}
\usepackage{graphicx}
\usepackage{latexsym}
\usepackage{bm}
\usepackage[colorlinks,citecolor=blue, linkcolor=blue,hyperindex,CJKbookmarks,dvipdfm]{hyperref}

\begin{document}

\title{Optomechanically induced nonreciprocity in a three-mode
optomechanical system}
\author{Xun-Wei Xu}
\email{davidxu0816@163.com}
\affiliation{Department of Applied Physics, East China Jiaotong University, Nanchang,
330013, China}
\author{L. N. Song}
\affiliation{Beijing Computational Science Research Center, Beijing 100193, China}
\author{Qiang Zheng}
\affiliation{School of Mathematics, Guizhou Normal University, Guiyang, 550001, China}
\affiliation{Beijing Computational Science Research Center, Beijing 100193, China}
\author{Z. H. Wang}
\affiliation{Center for Quantum Sciences and School of Physics, Northeast Normal
University, Changchun 130024, China}
\affiliation{Beijing Computational Science Research Center, Beijing 100193, China}
\author{Yong Li}
\email{liyong@csrc.ac.cn}
\affiliation{Beijing Computational Science Research Center, Beijing 100193, China}
\date{\today }

\begin{abstract}
We propose to create optical nonreciprocity in a three-mode optomechanical
system comprising one mechanical and two optical modes, where the
mechanical mode is coupled with only one of the optical modes. The optical
nonreciprocal response of the system is based on the nonlinearity induced by
the optomechanical interaction. However, nonlinearity is a necessary but not
a sufficient condition for observing nonreciprocity. Another necessary
condition for nonreciprocal
response of the system to a classical driving field is demonstrated
analytically. The effects of the parameters on the nonreciprocal response of
the system are discussed numerically. The three-mode optomechanical system
provides a platform to realize nonreciprocity for strong optical signal
fields.
\end{abstract}

\maketitle




\section{Introduction}

Optomechanical system with parameter coupling between optical and mechanical
modes provides us a perfect platform for various classical and quantum
information processing applications~\cite%
{KippenbergSci08,MarquardtPhy09,AspelmeyerPT12,AspelmeyerARX13,MetcalfeAPR14,YLLiuCPB18}%
. As an important application, the design of nonreciprocal devices, that
allow signals transmitting in one direction while blocking those propagating
in the opposite one, based on optomechanical interaction has attracted
significant interest in the past few years~\cite{VerhagenNPy17,SounasNPo17}.
A number of designs based on diverse mechanisms are proposed to demonstrate
nonreciprocity in optomechanical systems. As a non-magnetic scheme,
optomechanically induced nonreciprocity makes the all-optical controllable
isolators and circulators integrated on chip easily.

Optomechanically induced transparency (OMIT) is a phenomenon that a tunable
transparent window for a weak probe is induced by the optomechanical
coupling with a strong driving field injected into the red-detuned sideband
of an optomechanical system~\cite%
{AgarwalPRA10,WeisScience10,QLinNPo10,SafaviNaeiniNat11,TeufelNat11}. OMIT
in microring resonators provides us a practical way to achieve optical
nonreciprocity~\cite{HafeziOE12}. The symmetry for the transmission of weak
signals is broken in the presence of a strong driving field. Specifically, a
strong driving field in the red sideband of the system induces a transparent
transmission window for signals propagating in the same direction as the
driving field, but not in the opposite one, leading to optical isolation,
which has already been observed in experiments~\cite{ZShenNPo16,RuesinkNC16}%
. Similar to the OMIT, stimulated Brillouin scattering (SBS) between two
optical modes by a travelling acoustic mode can also induce transparency
when they satisfy both the energy and momentum conservations, and has been
used successfully for the generation of nonreciprocity in optomechanical
systems~\cite{CHDongNC15,JKimNPy15}.

Besides the optomechanical nonreciprocity in degenerate whispering-gallery
modes with inherent non-trivial topology, the nonreciprocity via synthetic
magnetism and reservoir engineering has been proposed theoretically~\cite%
{XWXuPRA15,MetelmannPRX15,XWXuPRA16,LTianPRA17,GLiPRA18,CJiangPRA18} and
realized experimently in an optomechanical circuit~\cite%
{KFangNPy17,BernierNC17,PetersonPRX17,BarzanjehNC17}. The time-reversal
symmetry is broken by a synthetic magnetic flux by phase-correlated driving
fields, which may enhance the photonic transport in one direction for
constructive quantum interference but suppress it in the reversal direction
due to destructive quantum interference.

Another approach to create nonreciprocity in the optomechanical system is by
using the Kerr-type nonlinear interaction induced by the mechanical mode. As
early as in 2009, the nonreciprocal optical transmission spectrum has been
demonstrated in a Fabry-Perot cavity with one movable mirror by the
asymmetry of the radiation pressure on the movable mirror for forward and
backward incident light~\cite{ManipatruniPRL09}. Moreover, the optical
nonreciprocity was also proposed in optomechanical systems with asymmetric
structures~\cite{ZWangSR15,HQiuOE17}.

However, there are substantial differences between the nonreciprocity based
on nonlinearity and the ones based on the other mechanisms mentioned above
(e.g., OMIT, SBS, and synthetic magnetic flux). For the nonreciprocity based
on nonlinearity, there is only one light beam injected into the
optomechanical system and the structure is nonreciprocal for this light,
which is usually pretty strong. Instead, the nonreciprocity based on the
other mechanisms is nonreciprocal for a weak light beam (signal field) in
the presence of a strong light beam (control field). When the signal field
is strong, the nonreciprocity based on nonlinearity would be a better choice.

In this paper, we propose the creation of optical nonreciprocity in a
three-mode optomechanical system based on the optomechanical induced nonlinearity~\cite{ManipatruniPRL09,ZWangSR15,HQiuOE17}.
Different from the previous works~\cite{ManipatruniPRL09,ZWangSR15,HQiuOE17}, here the setup consists of one mechanical and two optical (cavity) modes, and the
mechanical mode is coupled with only one of the optical modes.
Interestingly, we find that the nonlinearity is a necessary but not a
sufficient condition for observing nonreciprocity. Another condition is necessary for creating the
nonreciprocal response of the system to the classical driving field. The
effects of the parameters on the nonreciprocal response of the system are
discussed numerically.

The remainder of this paper is organized as follows. In Sec.~II, we
introduce the theoretical model of three-mode optomechanical system, and
derive the nonlinear equations for the output fields. In Sec.~III, the
nonreciprocal responses of the system to the classical driving field are
discussed numerically. Finally, a summary is given in Sec.~IV.

\section{Theoretical model}

\begin{figure}[tbp]
\includegraphics[bb=85 283 535 686, width=8.5 cm, clip]{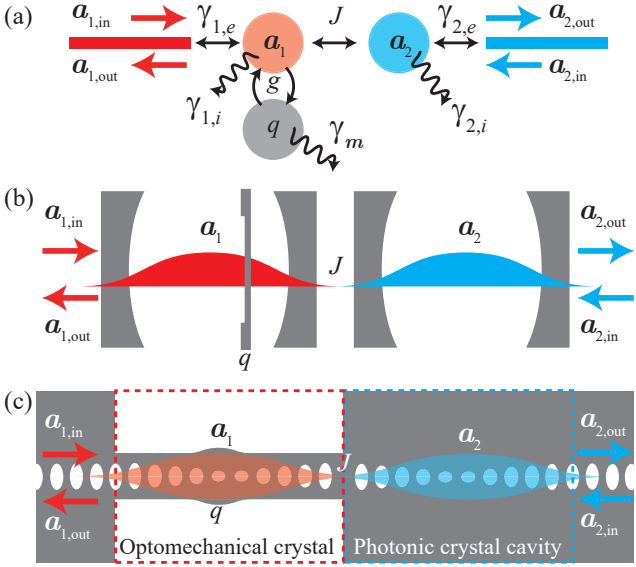}
\caption{(Color online) (a) Scheme of a three-mode optomechanical system
consisting of two linearly coupled optical modes ($a_{1}$ and $a_{2}$) and a
mechanical mode ($q$) coupled to one of the optical modes via optomechanical
interaction. (b) and (c) Example implementations of the three-mode
optomechanical system: (b) a membrane in the middle of one of the two
coupled optical cavities~\protect\cite%
{ThompsonNat08,Flowers-JacobsAPL12,HXuNat16,HXuNC17}, and (d) an
optomechanical crystal coupled to the photonic crystal cavity~\protect\cite%
{ParaisoPRX15,KFangNPy17}.}
\label{fig1}
\end{figure}

We consider an optomechanical system consisting of two linearly coupled
optical modes ($a_{1}$ and $a_{2}$, with frequencies $\omega _{1}$ and $%
\omega _{2}$) and one of them is coupled to a mechanical mode ($p$ and $q$,
with vibrational frequency $\omega _{m}$) through optomechanical
interaction, as shown in Fig.~\ref{fig1}(a), that is described by the
Hamiltonian ($\hbar =1$)
\begin{eqnarray}
H &=&\omega _{1}a_{1}^{\dag }a_{1}+\omega _{2}a_{2}^{\dag }a_{2}+\frac{1}{2}%
\omega _{m}\left( q^{2}+p^{2}\right)  \notag \\
&&+J\left( a_{1}^{\dag }a_{2}+a_{2}^{\dag }a_{1}\right) +ga_{1}^{\dag }a_{1}q
\notag \\
&&+i\sqrt{\eta _{1}\gamma _{1}}\left( a_{1}^{\dag }a_{1,\mathrm{in}%
}e^{-i\omega _{d}t}-a_{1}a_{1,\mathrm{in}}^{\dag }e^{i\omega _{d}t}\right)
\notag \\
&&+i\sqrt{\eta _{2}\gamma _{2}}\left( a_{2}^{\dag }a_{2,\mathrm{in}%
}e^{-i\omega _{d}t}-a_{2}a_{2,\mathrm{in}}^{\dag }e^{i\omega _{d}t}\right) .
\end{eqnarray}%
Here $J$ is the coupling strength between the two optical modes and $g$
denotes the single-photon optomechanical coupling strength. We assume that $%
\gamma _{j}=\gamma _{j,i}+\gamma _{j,e}$ ($j=1,2$) is the total decay rate
of the optical mode $a_{j}$ with the cavity coupling parameter $\eta
_{j}=\gamma _{j,e}/\left( \gamma _{j,i}+\gamma _{j,e}\right) $, where $%
\gamma _{j,i}$ denotes the intrinsec decay rate and $\gamma _{j,e}$ the
external decay rate (i.e. waveguide-to-cavity coupling). $a_{j,\mathrm{in}}$
and $\omega _{d}$ are the driving amplitude and frequency at the input of
the optical mode $a_{j} $ via the waveguide-to-cavity coupling $\gamma
_{j,e}=\eta _{j}\gamma _{j}$. Such a system can be implemented in two
linearly coupled Fabry-P\'{e}rot cavities with a membrane in the node of one
of the optical cavities [see Fig.~\ref{fig1}(b)]~\cite%
{ThompsonNat08,Flowers-JacobsAPL12,HXuNat16,HXuNC17}, and an optomechanical
crystal coupled to the photonic crystal cavity [see Fig.~\ref{fig1}(c)]~\cite%
{ParaisoPRX15,KFangNPy17}.

In the rotating frame of the driving frequency $\omega _{d}$, the dynamics
of the system is described by the quantum Langevin equations (QLEs)
\begin{eqnarray}
\frac{d}{dt}a_{1} &=&-\left( \frac{\gamma _{1}}{2}+i\Delta _{1}\right)
a_{1}-igqa_{1}-iJa_{2}  \notag \\
&&+\sqrt{\eta _{1}\gamma _{1}}a_{1,\mathrm{in}}+\sqrt{\left( 1-\eta
_{1}\right) \gamma _{1}}a_{1,\mathrm{vac}},
\end{eqnarray}%
\begin{eqnarray}
\frac{d}{dt}a_{2} &=&-\left( \frac{\gamma _{2}}{2}+i\Delta _{2}\right)
a_{2}-iJa_{1}  \notag \\
&&+\sqrt{\eta _{2}\gamma _{2}}a_{2,\mathrm{in}}+\sqrt{\left( 1-\eta
_{2}\right) \gamma _{2}}a_{2,\mathrm{vac}},
\end{eqnarray}%
\begin{equation}
\frac{d}{dt}q=\omega _{m}p,
\end{equation}%
\begin{equation}
\frac{d}{dt}p=-\omega _{m}q-ga_{1}^{\dag }a_{1}-\gamma _{m}p+\xi ,
\end{equation}%
where $\Delta _{j}=\omega _{j}-\omega _{d}$ is the detuning parameter, $%
\gamma _{m}$ is the decay rate of the mechanical mode; $a_{1,\mathrm{vac}}$,
$a_{2,\mathrm{vac}}$, and $\xi $ are the input quantum noise operators with
zero mean values, $\left\langle a_{1,\mathrm{vac}}\right\rangle =0$, $%
\left\langle a_{2,\mathrm{vac}}\right\rangle =0$, $\left\langle \xi
\right\rangle =0$.

Each operator of the optical and mechanical modes can be split into a
classical mean value and fluctuation: $a_{j}=\alpha _{j}+\delta a_{j}$, $a_{j%
\mathrm{,in}}=\alpha _{j\mathrm{,in}}+\delta a_{j,\mathrm{in}}$, $q=%
\overline{q}+\delta q$, $p=\overline{p}+\delta p$, with the ansatz $\alpha
_{j}=\left\langle a_{j}\right\rangle $, $\alpha _{j,\mathrm{in}%
}=\left\langle a_{j,\mathrm{in}}\right\rangle $, $\overline{q}=\left\langle
q\right\rangle $, $\overline{p}=\left\langle p\right\rangle $. Under the
condition of strong driving and weak optomechanical coupling, i.e., $\sqrt{%
\eta _{j}\gamma _{j}}|\alpha _{j,\mathrm{in}}|\gg \gamma_{j} \gg g$, the
classical mean values and quantum fluctuations can be treated separately. In
the steady state for $d\alpha _{i}/dt=0$ and $d\overline{p}/dt=d\overline{q}%
/dt=0$, the mean values ($\alpha _{i}$, $\overline{q}$) are given by
\begin{equation}  \label{eq6}
0=-\left( \frac{\gamma _{1}}{2}+i\Delta _{1}\right) \alpha _{1}-ig\overline{q%
}\alpha _{1}-iJ\alpha _{2}+\sqrt{\eta _{1}\gamma _{1}}\alpha _{1,\mathrm{in}%
},
\end{equation}%
\begin{equation}  \label{eq7}
0=-\left( \frac{\gamma _{2}}{2}+i\Delta _{2}\right) \alpha _{2}-iJ\alpha
_{1}+\sqrt{\eta _{2}\gamma _{2}}\alpha _{2,\mathrm{in}},
\end{equation}%
\begin{equation}
\omega _{m}\overline{q}=-g\left\vert \alpha _{1}\right\vert ^{2},
\end{equation}
where the mean-field approximation (e.g., $\left\langle qa_{1}\right\rangle
\approx \left\langle q\right\rangle \left\langle a_{1}\right\rangle $) is
used in the derivation. By introducing the vectors $f=\left(
\begin{array}{cccccc}
\delta a_{1} & \delta a_{1}^{\dag } & \delta a_{2} & \delta a_{2}^{\dag } &
\delta q & \delta p%
\end{array}%
\right) ^{T}$ and $\zeta =\left(
\begin{array}{cccccc}
\delta A_{1,\mathrm{in}} & \delta A_{1,\mathrm{in}}^{\dag } & \delta A_{2,%
\mathrm{in}} & \delta A_{2,\mathrm{in}}^{\dag } & 0 & \xi%
\end{array}%
\right) $, with $\delta A_{j,\mathrm{in}}=\sqrt{\eta _{j}\gamma _{j}}\delta
a_{j,\mathrm{in}}+\sqrt{\left( 1-\eta _{j}\right) \gamma _{j}}a_{j,\mathrm{%
vac}}$, the linearized QLEs for the fluctuation operators of the optical and
mechanical modes can be given in the matrix form as
\begin{equation}
\frac{d}{dt}f=Af+\zeta ,
\end{equation}%
where the coefficient matrix%
\begin{widetext}
\begin{equation}
A=\left(
\begin{array}{cccccc}
-\left( \frac{\gamma _{1}}{2}+i\Delta _{1}^{\prime }\right)  & 0 & -iJ & 0 &
-ig\alpha _{1} & 0 \\
0 & -\left( \frac{\gamma _{1}}{2}-i\Delta _{1}^{\prime }\right)  & 0 & +iJ &
+ig\alpha _{1}^{\ast } & 0 \\
-iJ & 0 & -\left( \frac{\gamma _{2}}{2}+i\Delta _{2}\right)  & 0 & 0 & 0 \\
0 & +iJ & 0 & -\left( \frac{\gamma _{2}}{2}-i\Delta _{2}\right)  & 0 & 0 \\
0 & 0 & 0 & 0 & 0 & \omega _{m} \\
-g\alpha _{1}^{\ast } & -g\alpha _{1} & 0 & 0 & -\omega _{m} & -\gamma _{m}%
\end{array}%
\right)
\end{equation}%
\end{widetext}
with the effective detuning $\Delta _{1}^{\prime }=\Delta _{1}+g\overline{q}$%
. The stability conditions for the system require that the real parts of all
the eigenvalues of matrix $A$ are negative, which can be given analytically
by using the Routh-Hurwitz criterion~\cite%
{DeJesusPRA87,Gradshteyn,PaternostroNJP06}. However, the analytical
conditions are too cumbersome to be given here. In the following, we will
check the stability of the system by calculating the eigenvalues of matrix $A
$ numerically.

In this paper, we focus on the mean response of the system to the classical
driving field. To discuss the difference of the mean response of the system
to the classical driving field input from different optical modes, we will
show the equations of the output fields when a classical field input from
the optical mode $a_{1}$ by setting $\alpha _{1,\mathrm{in}}=s_{\mathrm{in}}$
and $\alpha _{2,\mathrm{in}}=0$, or input from the optical mode $a_{2}$ by
setting $\alpha _{1,\mathrm{in}}=0$ and $\alpha _{2,\mathrm{in}}=s_{\mathrm{%
in}}$, respectively. Here, $s_{\mathrm{in}}=\sqrt{p_{\mathrm{in}}/(\hbar
\omega _{d})}$ is the amplitude of the driving field with pump power $p_{%
\mathrm{in}}$.

For the case of the classical field input from the optical mode $a_{1}$,
i.e., $\alpha _{1,\mathrm{in}}=s_{\mathrm{in}}$ and $\alpha _{2,\mathrm{in}%
}=0$, we will derive the equation for the output field $\alpha _{2,\mathrm{%
out}}$ from the optical mode $a_{2}$. Eqs.~(\ref{eq6}) and (\ref{eq7}) are
rewritten with $\overline{q}=-g\left\vert \alpha _{1}\right\vert ^{2}/\omega
_{m}$ as
\begin{equation}  \label{eq11}
0=-\left( \frac{\gamma _{1}}{2}+i\Delta _{1}\right) \alpha _{1}-iU\left\vert
\alpha _{1}\right\vert ^{2}\alpha _{1}-iJ\alpha _{2}+\sqrt{\eta _{1}\gamma
_{1}}s_{\mathrm{in}},
\end{equation}%
\begin{equation}  \label{eq12}
0=-\left( \frac{\gamma _{2}}{2}+i\Delta _{2}\right) \alpha _{2}-iJ\alpha
_{1},
\end{equation}%
where $U\equiv -g^{2}/\omega _{m}$ is the nonlinear interaction induced by
the optomechanical interaction. Using Eq.~(\ref{eq12}), one has
\begin{equation}  \label{eq12prime}
\alpha _{2}=\frac{-i2J}{\gamma _{2}+i2\Delta _{2}}\alpha _{1}.
\end{equation}%
Substituting it into the standard input-output relation~\cite{GardinerPRA85}
\begin{equation}  \label{eq14}
\alpha _{2,\mathrm{out}}=\sqrt{\eta _{2}\gamma _{2}}\alpha _{2}=\frac{-i2J%
\sqrt{\eta _{2}\gamma _{2}}}{\gamma _{2}+i2\Delta _{2}}\alpha _{1},
\end{equation}%
and Eq.~(\ref{eq11}), the equation for the output field $\alpha _{2,\mathrm{%
out}}$ can be written as
\begin{equation}  \label{eq15}
0=-\left( \frac{\Gamma }{2}+i\overline{\Delta }\right) \alpha _{2,\mathrm{out%
}}-iU_{21}\left\vert \alpha _{2,\mathrm{out}}\right\vert ^{2}\alpha _{2,%
\mathrm{out}}+\varepsilon _{\mathrm{eff}}
\end{equation}%
with the effective damping rate $\Gamma$, detuning $\overline{\Delta }$,
nonlinear interaction strength $U_{21}$, and driving amplitude $\varepsilon
_{\mathrm{eff}}$ given by%
\begin{equation}
\Gamma\equiv \gamma _{1}+\frac{4\gamma _{2}J^{2}}{\gamma _{2}^{2}+4\Delta
_{2}^{2}},
\end{equation}%
\begin{equation}
\overline{\Delta }\equiv \Delta _{1}-\frac{4J^{2}\Delta _{2}}{\gamma
_{2}^{2}+4\Delta _{2}^{2}},
\end{equation}%
\begin{equation}  \label{eq18}
U_{21}\equiv \frac{\gamma _{2}^{2}+4\Delta _{2}^{2}}{4\eta _{2}\gamma
_{2}J^{2}}U,
\end{equation}%
\begin{equation}
\varepsilon _{\mathrm{eff}}\equiv -\frac{i2J\sqrt{\eta _{1}\gamma _{1}\eta
_{2}\gamma _{2}}}{\gamma _{2}+i2\Delta _{2}}s_{\mathrm{in}}.
\end{equation}

When the classical field is input from the optical mode $a_{2}$, i.e., $%
\alpha _{1,\mathrm{in}}=0$ and $\alpha _{2,\mathrm{in}}=s_{\mathrm{in}}$, we
will consider the output field $\alpha _{1,\mathrm{out}}$ from the optical
mode $a_{1}$. In this case, Eqs.~(\ref{eq6}) and (\ref{eq7}) are rewritten
as
\begin{equation}  \label{eq20}
0=-\left( \frac{\gamma _{1}}{2}+i\Delta _{1}\right) \alpha _{1}-iU\left\vert
\alpha _{1}\right\vert ^{2}\alpha _{1}-iJ\alpha _{2},
\end{equation}%
\begin{equation}  \label{eq21}
0=-\left( \frac{\gamma _{2}}{2}+i\Delta _{2}\right) \alpha _{2}-iJ\alpha
_{1}+\sqrt{\eta _{2}\gamma _{2}}s_{\mathrm{in}}.
\end{equation}%
From Eq.~(\ref{eq21}) one has
\begin{equation}
\alpha _{2}=\frac{-i2J}{\gamma _{2}+i2\Delta _{2}}\alpha _{1}+\frac{2\sqrt{%
\eta _{2}\gamma _{2}}}{\gamma _{2}+i2\Delta _{2}}s_{\mathrm{in}},
\end{equation}%
which together with Eq.~(\ref{eq20}) and the input-output relation~\cite%
{GardinerPRA85}
\begin{equation}
\alpha _{1,\mathrm{out}}=\sqrt{\eta _{1}\gamma _{1}}\alpha _{1}
\end{equation}%
provide the following equation for the output field $\alpha _{1,\mathrm{out}%
} $ as
\begin{equation}  \label{eq24}
0=-\left( \frac{\Gamma }{2}+i\overline{\Delta }\right) \alpha _{1,\mathrm{out%
}}-iU_{12}\left\vert \alpha _{1,\mathrm{out}}\right\vert ^{2}\alpha _{1,%
\mathrm{out}}+\varepsilon _{\mathrm{eff}}
\end{equation}%
with the effective nonlinear interaction strength
\begin{equation}  \label{eq25}
U_{12}\equiv \frac{U}{\eta _{1}\gamma _{1}}.
\end{equation}

In order to find the output field $\left\vert \alpha _{i,\mathrm{out}%
}\right\vert ^{2}$ ($i=1,2$), Eqs.~(\ref{eq15}) and (\ref{eq24}) can be
rewritten in a unified form as%
\begin{equation}
\left( \frac{\Gamma ^{2}}{4}+\overline{\Delta }^{2}\right) \left\vert \alpha
_{i,\mathrm{out}}\right\vert ^{2}+2\overline{\Delta }U_{\mathrm{eff}%
}\left\vert \alpha _{i,\mathrm{out}}\right\vert ^{4}+U_{\mathrm{eff}%
}^{2}\left\vert \alpha _{i,\mathrm{out}}\right\vert ^{6}=\left\vert
\varepsilon _{\mathrm{eff}}\right\vert ^{2},  \label{eq27}
\end{equation}%
where $U_{\mathrm{eff}}=U_{\mathrm{12}}$ for $\left\vert \alpha _{1,\mathrm{%
out}}\right\vert ^{2}$ and $U_{\mathrm{eff}}=U_{\mathrm{21}}$ for $%
\left\vert \alpha _{2,\mathrm{out}}\right\vert ^{2}$. The output field $%
\left\vert \alpha _{i,\mathrm{out}}\right\vert ^{2}$ is found to have three
possible real values, the so-called optical bistability, under some
conditions. One way to get the bistability condition is to take a derivative
of Eq.~(\ref{eq27}) with respect to $\left\vert \alpha _{i,\mathrm{out}%
}\right\vert ^{2}$, then we have
\begin{equation}
\left( \frac{\Gamma ^{2}}{4}+\overline{\Delta }^{2}\right) +4\overline{%
\Delta }U_{\mathrm{eff}}\left\vert \alpha _{i,\mathrm{out}}\right\vert
^{2}+3U_{\mathrm{eff}}^{2}\left\vert \alpha _{i,\mathrm{out}}\right\vert
^{4}=0.
\end{equation}%
Thus the value of $\left\vert \alpha _{i,\mathrm{out}}\right\vert ^{2}$
corresponding to the bistability turning point is the two roots of the above
equation,
\begin{equation}
\left\vert \alpha _{i,\mathrm{out}}\right\vert _{\pm }^{2}=\frac{-4\overline{%
\Delta }\mp \sqrt{4\overline{\Delta }^{2}-3\Gamma ^{2}}}{6U_{\mathrm{eff}}}>0
\end{equation}%
with the bistability condition (positive values of the discriminant)%
\begin{equation}
\overline{\Delta }>\frac{\sqrt{3}}{2}\Gamma .
\end{equation}%
It is well known that the intermediate value in the bistability region is
unstable for it corresponds to the maximum (not minimum) point in the
effective potential energy. Different from the Kerr medium, where the upper
and lower values are always stable, the other type of instability may appear
in the upper value of an optomechanical system due to the heating of the
mechanical mode, which may result in the negative effective damping of the mechanical mode in the blue-detuned regime or strong driving condition~\cite%
{CarmonPRL05,KippenbergPRL05,MarquardtPRL06,LudwigNJP08,AldanaPRA13}.

The transmission coefficient in the direction from the optical mode $a_{1}$
to the optical mode $a_{2}$ is defined by
\begin{equation}
T_{21}\equiv \left\vert \frac{\alpha _{2,\mathrm{out}}}{\alpha _{1,\mathrm{in%
}}}\right\vert ^{2},
\end{equation}%
and the transmission coefficient in the opposite direction is defined by
\begin{equation}
T_{12}\equiv \left\vert \frac{\alpha _{1,\mathrm{out}}}{\alpha _{2,\mathrm{in%
}}}\right\vert ^{2}.
\end{equation}%
The efficiency for the nonreciprocity transmission can be described by the
isolation
\begin{equation}
I=\frac{T_{21}}{T_{12}}.
\end{equation}%
From the nonlinear equations for $\alpha _{2,\mathrm{out}}$ [Eq.~(\ref{eq15}%
)] and $\alpha _{1,\mathrm{out}}$ [Eq.~(\ref{eq24})], it is interesting to
note that, when
\begin{equation}
\sqrt{\eta _{1}\gamma _{1}}=\left\vert \frac{-i2J}{\gamma _{2}+i2\Delta _{2}}%
\right\vert \sqrt{\eta _{2}\gamma _{2}},  \label{eq28}
\end{equation}%
the reciprocity will be observed for $U_{21}=U_{12}$, even when $U_{21}\neq 0
$. $\sqrt{\eta _{1}\gamma _{1}}$ is the coupling rate of the optical mode $%
a_{1,\mathrm{in}}$ to $a_{1}$, and $\left\vert -i2J/(\gamma _{2}+i2\Delta
_{2})\right\vert \sqrt{\eta _{2}\gamma _{2}}$ is the effective coupling rate
of the optical mode $a_{1}$ to $a_{2,\mathrm{out}} $ through the optical
mode $a_{2}$ [see Eq.~(\ref{eq14})]. We refer to Eq.~(\ref{eq28}) as the
impedance-matching condition for these two coupling rates are equal. On the
contrary, if the impedance-matching condition is broken, we will have $%
U_{21}\neq U_{12}$, i.e., the effective nonlinearity induced by the
optomechanical interaction is different for photons transport in different
directions, and as a result the optical nonreciprocity appears in the
three-mode optomechanical system, i.e., $T_{21}\neq T_{12}$. That is to say
nonlinearity is a necessary but not a sufficient condition for observing
nonreciprocity, and breaking the impedance-matching condition is another
necessary condition for nonreciprocal response of the system.

\section{Nonreciprocal transmission}

\begin{figure}[tbp]
\includegraphics[bb=111 221 530 557, width=8.5 cm, clip]{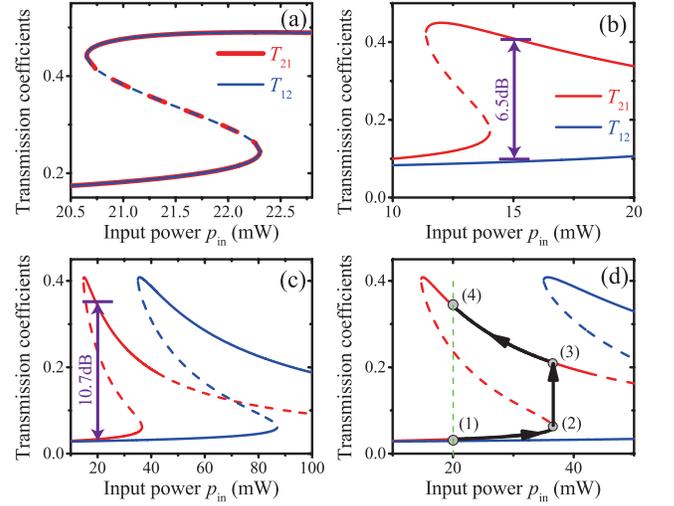}
\caption{(Color online) Transmission coefficients $T_{21}$ (red) and $T_{12}$
(blue) as a function of the input power $p_{\mathrm{in}}$ ($\mathrm{mW}$).
(a) $\Delta_{1}/2\protect\pi=2$ GHz, $\Delta_{2}=0$, $J/2\protect\pi=0.5$
GHz; (b) $\Delta_{1}/2\protect\pi=\Delta_{2}/2\protect\pi=4$ GHz, $J/2%
\protect\pi=3$ GHz; (c) and (d) $\Delta_{1}/2\protect\pi=\Delta_{2}/2\protect%
\pi=4.6$ GHz, $J/2\protect\pi=3$ GHz. The dashed curves correspond to the
unstable regimes for the eigenvalues of the matrix $A$ containing positive
real parts. The other parameters are $\protect\omega_d/2\protect\pi= 200$
THz, $\protect\gamma_{1}/2\protect\pi=\protect\gamma_{2}/2\protect\pi=1$
GHz, $\protect\eta_{1}=\protect\eta_{2}=0.7$, $\protect\omega_m/2\protect\pi%
=6$ GHz, $\protect\gamma_m/2\protect\pi=5$ MHz, and $g/2\protect\pi=0.8$
MHz. }
\label{fig2}
\end{figure}

\begin{figure}[tbp]
\includegraphics[bb=87 202 526 555, width=8.5 cm, clip]{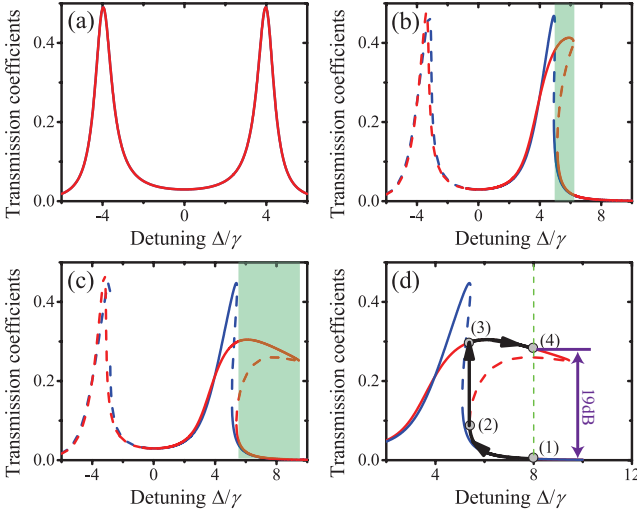}
\caption{(Color online) Transmission coefficients $T_{21}$ (red) and $T_{12}$
(blue) as a function of the detuning $\Delta/\protect\gamma$ ($%
\Delta\equiv\Delta_1=\Delta_2, \protect\gamma\equiv\protect\gamma_1=\protect%
\gamma_2$). (a) $p_{\mathrm{in}}=0.1$ mW; (b) $p_{\mathrm{in}}=20$ mW; (c)
and (d) $p_{\mathrm{in}}=30$ mW. The other parameters are $\protect\omega_d/2%
\protect\pi\approx 200$ THz, $\protect\gamma_{1}/2\protect\pi=\protect\gamma%
_{2}/2\protect\pi=1$ GHz, $J/2\protect\pi=4$ GHz, $\protect\eta_{1}=\protect%
\eta_{2}=0.7$, $\protect\omega_m/2\protect\pi=6$ GHz, $\protect\gamma_m/2%
\protect\pi=5$ MHz, and $g/2\protect\pi=0.8$ MHz. The region for notable
nonreciprocity is marked out with green colour.}
\label{fig3}
\end{figure}

In this section, we will discuss the effects of the parameters on the
nonreciprocal transmission of light input from different optical modes by
solving the equations numerically. For numerical simulation, we use
parameters based on a recent experiment on the observation of nonreciprocity
in an optomechanical crystal~\cite{KFangNPy17}: $\omega_d/2\pi\approx 200$
THz, $\gamma_{1}/2\pi=\gamma_{2}/2\pi=1$ GHz, $\eta_{1}=\eta_{2}=0.7$, $%
\omega_m/2\pi=6$ GHz, $\gamma_m/2\pi=5$ MHz, and $g/2\pi=0.8$ MHz.

In Fig.~\ref{fig2}, we show the transmission coefficients $T_{21}$ (red) and
$T_{12}$ (blue) as a function of the input power $p_{\mathrm{in}}$. First of
all, let us check the necessary condition for nonreciprocity numerically.
Although the system works in the bistable regime, as shown in Fig.~\ref{fig2}%
(a), the nonreciprocity can not be observed, i.e., $T_{12}=T_{21}$, when the
impedance-matching condition given by Eq.~(\ref{eq28}) is satisfied.
Conversely, the nonreciprocity appears when the impedance-matching condition
is broken, as shown in Fig.~\ref{fig2}(b). The related isolation is $%
I\approx 6.5$ dB for input power $p_{\mathrm{in}}=15 $ mW, with the
detunings $\Delta_{1}/2\pi=\Delta_{2}/2\pi=4$ GHz and coupling strength $%
J/2\pi=3$ GHz. The isolation can be improved to be $I\approx 10.7$ dB for
input power $p_{\mathrm{in}}=20 $ mW when the detunings $\Delta_{1}/2\pi=%
\Delta_{2}/2\pi=4.6$ GHz, as shown in Fig.~\ref{fig2}(c). In order to make
the system work in the upper branch, see Fig.~\ref{fig2}(d), we can increase
the input power (1) to the critical power (2) [or (3)] for optical
bistability, and then reduce the input power to the working power in the
upper branch (4).

We note that it is possible to achieve optical isolation with high
transmission in the direction from $a_{2}$ to $a_{1}$, as shown in Fig.~\ref%
{fig2}(c) around $p_{\mathrm{in}}=40$ mW. However, the isolation in this
direction is much lower than the isolation can be achieved in the opposite
direction. Moreover, the require power is much higher, and worse yet, the
high power with $p_{\mathrm{in}}>44$ mW will cause instability for field
input from $a_{1}$, as shown in Fig.~\ref{fig2}(c). So we only consider the
optical isolation with high transmission from $a_{1}$ to $a_{2}$ in the
following.

The transmission coefficients $T_{21}$ (red) and $T_{12}$ (blue) plotted as
a function of the detuning $\Delta/\gamma$ ($\Delta\equiv\Delta_1=\Delta_2,
\gamma\equiv\gamma_1=\gamma_2$) are shown in Fig.~\ref{fig3} for different
input powers. When the input power is low, e.g., $p_{\mathrm{in}}=0.1$ mW,
there are peaks around the two resonant frequencies with the detuning $%
\Delta\approx\pm J$, and the nonreciprocity effect is not pronounced. As the
input power increases, the optical bistability occurs and the nonreciprocity
effect becomes visible with the detuning $\Delta>J$ [see Figs.~\ref{fig3}(b)
and \ref{fig3}(c)]. In order to make the system work in the upper branch,
see Fig.~\ref{fig3}(d), we can tune the frequency of the input field from
(1) to the critical detuning (2) [or (3)] for the optical bistability, and
then increase the frequency of the input field to the working frequency in
the upper branch (4). In the case with $\Delta=2J $, $p_{\mathrm{in}}=30$
mW, and $J=2\pi\times 6$ GHz, the system can work with the transmission
coefficients $T_{21}\approx 0.28$, $T_{12}\approx 0.00335$, and the
isolation $I\approx 19 $ dB.

It is worth mentioning that the optical multiple solutions may also occur
with the detuning satisfying $-J<\Delta<0$, but the system is unstable in
this regime, which are shown by dashed curves in Figs.~\ref{fig3}(b) and \ref%
{fig3}(c). This instability is induced by the heating of the mechanical mode
in the blue-detuned regime $\Delta<0$, for the effective damping of the
mechanical mode may become negative~\cite%
{CarmonPRL05,KippenbergPRL05,MarquardtPRL06,LudwigNJP08,AldanaPRA13}.

Different from the behavior of bistability in a standard optomechanical
system with one mechanical mode coupled to one optical mode~\cite%
{AldanaPRA13}, the peaks ``arch down'' as a function of detuning, as shown in
Fig.~\ref{fig3}(c).
This phenomenon is induced by the
detuning-dependent nonlinear interaction strength $U_{21}$ in Eq.~(\ref{eq18}%
). More specifically, the nonlinear interaction strength $U_{21}$ increases
with increasing detuning, and the transmission coefficient $T_{21}$ is
suppressed by the increasing nonlinear interaction strength $U_{21}$ with
increasing detuning. By contrast, there is no such suppression in $T_{12}$
for the constant nonlinear interaction strength $U_{12}$ in Eq.~(\ref{eq25}).

\begin{figure}[tbp]
\includegraphics[bb=98 387 516 557, width=8.5 cm, clip]{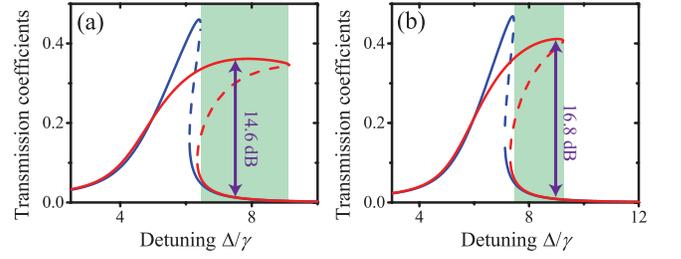}
\caption{(Color online) Transmission coefficients $T_{21}$ (red) and $T_{12}$
(blue) as a function of the detuning $\Delta/\protect\gamma$ ($%
\Delta\equiv\Delta_1=\Delta_2, \protect\gamma\equiv\protect\gamma_1=\protect%
\gamma_2$). (a) $J/2\protect\pi=5$ GHz; (b) $J/2\protect\pi=6$ GHz. The
other parameters are $\protect\omega_d/2\protect\pi\approx 200$ THz, $%
\protect\gamma_{1}/2\protect\pi=\protect\gamma_{2}/2\protect\pi=1$ GHz, $p_{%
\mathrm{in}}=30$ mW, $\protect\eta_{1}=\protect\eta_{2}=0.7$, $\protect\omega%
_m/2\protect\pi=6$ GHz, $\protect\gamma_m/2\protect\pi=5$ MHz, and $g/2%
\protect\pi=0.8$ MHz. The region for notable nonreciprocity is marked out
with green colour.}
\label{fig4}
\end{figure}

\begin{figure}[tbp]
\includegraphics[bb=72 383 502 557, width=8.5 cm, clip]{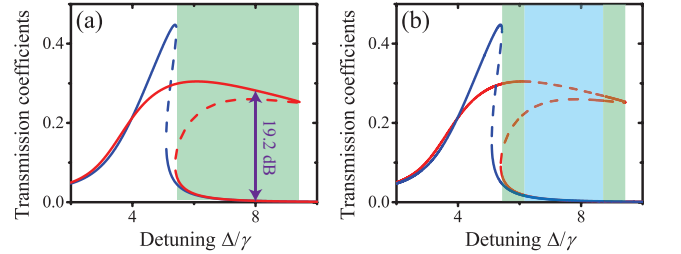}
\caption{(Color online) Transmission coefficients $T_{21}$ (red) and $T_{12}$
(blue) as a function of the detuning $\Delta/\protect\gamma$ ($%
\Delta\equiv\Delta_1=\Delta_2, \protect\gamma\equiv\protect\gamma_1=\protect%
\gamma_2$). (a) $\protect\omega_m/2\protect\pi=6$ GHz, $%
g/2\protect\pi=8$ MHz and $p_{\mathrm{in}}=0.3$ mW; (b) $\protect\omega_m/2%
\protect\pi=6/1.1$ GHz, $g/2\protect\pi=0.8$ MHz and $p_{\mathrm{in}}=30/1.1$
mW. The other parameters
are $\protect\omega_d/2\protect\pi\approx 200$ THz, $\protect\gamma_{1}/2%
\protect\pi=\protect\gamma_{2}/2\protect\pi=1$ GHz, $J/2\protect\pi=4$ GHz, $%
\protect\eta_{1}=\protect\eta_{2}=0.7$, and $\protect\gamma_m/2\protect\pi=5$
MHz. The region for notable nonreciprocity is marked out with green colour,
and the region for unstable upper branch is marked out with blue colour.}
\label{fig5}
\end{figure}

Now we will discuss the ideal range of parameters for notable nonreciprocity
with high transmission coefficients. Figure~\ref{fig4} presents the
transmission coefficients $T_{21}$ (red) and $T_{12}$ (blue) as a function
of the detuning $\Delta/\gamma$ for different coupling strength $J$. It is
clear that both the isolation $I $ and the transmission coefficient $T_{21}$
are improved with a larger coupling strength $J$. However, the frequency
range for the nonreciprocity (marked out with green colour) becomes
narrower. In some experiments~\cite{BPengNPy14,LChangNPo14}, the coupling
strength $J$ between the two optical cavities is tunable by adjusting the
gap between two cavities.

The optomechanical coupling strength $g$ and the frequency of the mechanical
mode $\omega_m$ are two critical parameters in optomechanical systems for
the nonlinear interaction strength $U=-g^{2}/\omega_m$. The nonlinear
interaction strength $U$ can be enhanced with a stronger optomechanical
coupling strength $g$ or lower mechanical frequency $\omega_m$. As shown in
Fig.~\ref{fig5}(a), if $g$ is enhanced by one order of
magnitude, the input power will be reduced by two orders of magnitude. That
means the optomechanically induced nonreciprocity can be achieved with a
much lower input power if the coupling strength $g$ can be enhanced in
experiments.

As shown in Fig.~\ref{fig5}(b), the input power for realizing
optomechanically induced nonreciprocity can also be reduced with lower
mechanical frequency $\omega_m$. However, the upper branch becomes unstable
(marked out with blue colour) due to the position fluctuations of the
mechanical mode for the effective damping of the mechanical mode may become
negative~\cite%
{CarmonPRL05,KippenbergPRL05,MarquardtPRL06,LudwigNJP08,AldanaPRA13}.
The unstable upper branch will also appear in the case of lower mechanical
frequency $\omega_m/k$ and smaller optomechanical coupling strength $g/\sqrt{%
k}$ ($k>1$), with the nonlinear
interaction strength $U=-g^{2}/\omega_m$ unchanged. Therefore, mechanical modes of
high frequency are conducive to the stability of the systems.

\section{Conclusions}

In summary, we have shown that the three-mode optomechanical system can be
used to realize the nonreciprocity for a strong light beam injected into the
system, due to the existence of the optomechanical nonlinearity. The
nonlinearity is a necessary, but not a sufficient condition for observing
nonreciprocal transport. Breaking of impedance-matching condition is another
pivotal condition for nonreciprocity. By adjusting the tunable parameters,
we can realize the nonreciprocity with both high isolation and high
transmission coefficients in the allowed transmission direction. Our work
paves the way towards the nonreciprocal transmission for strong optical
signals in optomechanical systems.
There are still two points that need be specified because they may restrict the applications of our proposal in some fields. First, the optical
isolation is highly sensitive to given parameters and particularly to the
power of the propagating light. Secondly, if
system works under bistable conditions, it has to be initialized by varying
the detuning or the strength every time that the optical isolation is
required. Therefore how to overcome these issues or even use them for useful applications need to be further studied in the future.

\vskip 2pc \leftline{\bf Acknowledgement}

We thank Dr. Cheng Jiang for helpful discussions. X.W.X. was supported by
the National Natural Science Foundation of China (NSFC) under Grant
No.~11604096. Z.H.W. was supported by the NSFC under Grant No.~11875011.
Y.L. acknowledges the support from the NSFC under Grant No.~11774024,
No.~11534002, and No.~U1530401.

\bibliographystyle{apsrev}
\bibliography{ref}

\end{document}